\documentclass[twocolumn,shortnote]{jpsj3}
\usepackage{xcolor}
\usepackage{txfonts}

\title{%
Projector Form of the Quantum Brachistochrone and Its Relation to
Two-Boundary Quantum Algorithm Design
}

\author{%
Masayuki \textsc{Ohzeki}$^{1,2,3,4}$\thanks{E-mail address: mohzeki@tohoku.ac.jp}
}

\inst{%
$^{1}$Graduate School of Information Sciences, Tohoku University, Sendai 980-8579, Japan\\
$^{2}$Department of Physics, Institute of Science Tokyo, Tokyo 152-8551, Japan\\
$^{3}$Research and Education Institute for Semiconductors and Informatics,
Kumamoto University, Kumamoto 860-8555, Japan\\
$^{4}$Sigma-i Co., Ltd., Tokyo 108-0075, Japan
}

\recdate{\today}

\abst{%
We clarify the relation between the time-optimal quantum evolution of
Carlini et al. and the coherent two-boundary optimal-control formulation
recently proposed for quantum algorithm design.  
}

\kword{%
quantum brachistochrone, quantum optimal control, quantum algorithms
}

\newcommand{\ket}[1]{|#1\rangle}
\newcommand{\bra}[1]{\langle #1|}
\newcommand{\braket}[2]{\langle #1|#2\rangle}

\providecommand{\Tr}{\mathop{\mathrm{Tr}}}
\newcommand{\ii}{\mathrm{i}}

\begin{document}
\maketitle

\noindent
Quantum optimal control admits two closely related but conceptually
different formulations.  
The first asks for the shortest time in which an initial pure state can be transformed into a final pure state under a fixed Hamiltonian resource.  This is the quantum brachistochrone problem studied
by Carlini, Hosoya, Koike, and Okudaira.\cite{Carlini2006}  
The second
fixes the endpoint time and minimizes the integrated quadratic effort of
the Hamiltonian.  This second viewpoint appears naturally in a coherent
quantum Schrodinger bridge formulation of quantum algorithm
design.\cite{Ohzeki2026,Ohzeki2026-2}
The purpose of this note is to make precise what is common and what is
different in these two viewpoints.  
In the simplest rank-one pure-state
setting with full Hermitian controllability and an isotropic Hamiltonian
norm, both formulations select the same Fubini--Study geodesic \cite{AnandanAharonov1990}.  
Their optimal Hamiltonians differ only by the choice of scale: Carlini et al. fix the available Hamiltonian norm and minimize time, whereas the bridge
formulation fixes time and minimizes quadratic action.  Outside this
special limit, however, the equivalence should not be assumed.

We also show that Carlini et al.'s Gram--Schmidt expression can be written
directly in terms of endpoint projectors.  This provides a useful bridge to
the quantum-computing interpretation: the commutator of boundary projectors
is the ideal continuous generator, and a quantum circuit is a synthesis or
approximation of this generator in the available control algebra.

\medskip
\noindent
Let $\ket{\psi(t)}$ be a normalized state evolving according to
\begin{equation}
  \frac{d}{dt}\ket{\psi(t)}=-\ii H(t)\ket{\psi(t)} ,
\end{equation}
where $\hbar=1$.  Carlini et al. formulate a time-minimization problem on
the projective Hilbert space.  Since the identity component of the
Hamiltonian changes only the global phase, the physically relevant part is
the traceless Hamiltonian
\begin{equation}
  \widetilde H=H-\frac{\Tr H}{n}I .
\end{equation}
For the isotropic constraint, the resource is fixed by
\begin{equation}
  \frac{1}{2}\Tr \widetilde H^2=\omega^2 .
  \label{eq:carlini_constraint}
\end{equation}

Choose the endpoint phase such that
\begin{equation}
  \braket{\psi_f}{\psi_i}=\cos\Theta,\qquad 0<\Theta<\frac{\pi}{2},
  \label{eq:geodesic_gauge}
\end{equation}
and define the normalized component of the final state orthogonal to the
initial state by
\begin{equation}
  \ket{\psi_f'}
  =
  \frac{\ket{\psi_f}-\cos\Theta\,\ket{\psi_i}}{\sin\Theta}.
  \label{eq:gs}
\end{equation}
The optimal traceless Hamiltonian obtained by Carlini et al. can then be
written as
\begin{equation}
  \widetilde H_C
  =
  \ii\omega
  \left(
    \ket{\psi_f'}\bra{\psi_i}
    -
    \ket{\psi_i}\bra{\psi_f'}
  \right).
  \label{eq:carlini_hamiltonian}
\end{equation}
The corresponding geodesic evolution is
\begin{equation}
  \ket{\psi(t)}
  =
  \cos(\omega t)\ket{\psi_i}
  +
  \sin(\omega t)\ket{\psi_f'} ,
  \label{eq:carlini_path}
\end{equation}
and the minimum time is
\begin{equation}
  T_{\min}=\frac{\Theta}{|\omega|}.
  \label{eq:carlini_time}
\end{equation}
The full Hamiltonian may still contain an arbitrary identity gauge term.
This freedom does not affect the projective trajectory.

The Gram--Schmidt basis in Eq.~(\ref{eq:gs}) appears because the two
endpoint states are generally not orthogonal.  It is therefore a coordinate
choice that turns the two-dimensional bridge plane into an ordinary
orthonormal plane in which the geodesic is a simple rotation.

\medskip
\noindent
In the coherent two-boundary formulation, the endpoint time $T$ is fixed
and one minimizes the quadratic Hamiltonian effort
\begin{equation}
  J[H]=\frac{1}{2}\int_0^T \Tr H(t)^2\,dt
  \label{eq:bridge_action}
\end{equation}
subject to the same projective boundary conditions.  For rank-one
endpoints, define
\begin{equation}
  P_i=\ket{\psi_i}\bra{\psi_i},\qquad
  P_f=\ket{\psi_f}\bra{\psi_f}.
\end{equation}
The full-control optimum is the constant geodesic Hamiltonian
\begin{equation}
  H_B
  =
  \frac{\Theta}{T\sin\Theta\cos\Theta}\,
  \ii [P_f,P_i],
  \label{eq:bridge_commutator}
\end{equation}
where
\begin{equation}
  \Theta=\arccos |\braket{\psi_f}{\psi_i}|.
\end{equation}
Equivalently, in the Gram--Schmidt basis of Eq.~(\ref{eq:gs}),
\begin{equation}
  H_B
  =
  \ii\frac{\Theta}{T}
  \left(
    \ket{\psi_f'}\bra{\psi_i}
    -
    \ket{\psi_i}\bra{\psi_f'}
  \right).
  \label{eq:bridge_gs}
\end{equation}
The minimum action is
\begin{equation}
  J_{\min}(T)=\frac{\Theta^2}{T}.
  \label{eq:bridge_min_action}
\end{equation}

Equation~(\ref{eq:bridge_commutator}) is the central simplification for
algorithm design.  The endpoint projectors determine the ideal continuous
direction.  
The remaining problem is to synthesize this direction using the
available controls, in a gate model, to approximate the same rotation by a finite circuit.

\medskip
\noindent
Carlini et al.'s expression can be written without explicitly introducing
the Gram--Schmidt vector.  Substituting
\begin{equation}
  \ket{\psi_f}
  =
  \cos\Theta\,\ket{\psi_i}
  +
  \sin\Theta\,\ket{\psi_f'}
  \label{eq:endpoint_decomposition}
\end{equation}
into the commutator of endpoint projectors gives
\begin{equation}
  [P_f,P_i]
  =
  \sin\Theta\cos\Theta
  \left(
    \ket{\psi_f'}\bra{\psi_i}
    -
    \ket{\psi_i}\bra{\psi_f'}
  \right).
  \label{eq:commutator_identity}
\end{equation}
Therefore, for nonorthogonal endpoint rays,
\begin{equation}
  \widetilde H_C
  =
  \omega\,
  \frac{\ii[P_f,P_i]}{\sin\Theta\cos\Theta}.
  \label{eq:carlini_projector}
\end{equation}
This is the same normalized direction as Eq.~(\ref{eq:bridge_commutator}),
with the replacement
\begin{equation}
  \omega=\frac{\Theta}{T}.
  \label{eq:scale_identification}
\end{equation}

The sign convention is important.  With the convention used here,
Carlini's orientation corresponds to $\ii[P_f,P_i]$.  Replacing it by
$\ii[P_i,P_f]$ reverses the direction of the projective motion, or
equivalently changes the sign of $\omega$.

\medskip
\noindent
The same projector formula contains the counterdiabatic term in the local
limit.  For $P(t)=\ket{n(t)}\bra{n(t)}$ and target $P(t+\delta t)$,
\begin{equation}
  \ii[P(t+\delta t),P(t)]
  =
  \ii[\dot P(t),P(t)]\delta t+O(\delta t^2),
  \label{eq:cd_limit}
\end{equation}
Thus $H_B(t,t+\delta t) \to \ii[\dot P(t),P(t)]$ because the nearby Fubini--Study angle is $v(t)\delta t+O(\delta t^2)$.  
The limit is Kato's parallel-transport  generator and Berry's rank-one counterdiabatic term.\cite{Kato,Berry}
Indeed, $\ii[\dot P,P]\ket n
=\ii(\ket{\dot n}-\braket{n}{\dot n}\ket n)$, and in the gauge
$\braket n{\dot n}=0$ this is
$H_{\rm CD}^{(n)}
=\ii(\ket{\dot n}\bra n-\ket n\bra{\dot n})$.  Thus the known local
counterdiabatic generator is the infinitesimal endpoint limit of the
finite projector bridge.

\medskip
\noindent
The two action principles are not identical before reduction.  Carlini et
al. minimize time under the norm constraint
Eq.~(\ref{eq:carlini_constraint}), whereas the bridge formulation minimizes
the quadratic action Eq.~(\ref{eq:bridge_action}) at fixed $T$.  Their
equivalence in the rank-one full-control case is the standard relation
between length and energy functionals on a Riemannian manifold.

Let $\gamma$ be the projective path and define its Fubini--Study length by
\begin{equation}
  L[\gamma]=\int ds .
\end{equation}
The fixed-time energy functional is
\begin{equation}
  E_T[\gamma]=\int_0^T \|\dot\gamma(t)\|_{\rm FS}^2\,dt .
\end{equation}
Cauchy's inequality gives
\begin{equation}
  E_T[\gamma]\geq \frac{L[\gamma]^2}{T},
  \label{eq:cs}
\end{equation}
with equality only for constant-speed motion.  Hence the fixed-time
quadratic-action problem selects a constant-speed shortest geodesic.  The
time-optimal problem selects the same geodesic traversed at the maximum
allowed speed.  For the endpoint distance $\Theta$, this gives
Eqs.~(\ref{eq:carlini_time}) and (\ref{eq:bridge_min_action}).

\medskip
\noindent
The equivalence described above is restricted.  It relies on rank-one
pure-state boundaries, full Hermitian controllability, and an isotropic
Hamiltonian metric.  If the allowed Hamiltonians are constrained to the
form
\begin{equation}
  H(t)=\sum_\mu u_\mu(t)A_\mu ,
\end{equation}
the ideal generator $\ii[P_f,P_i]$ need not lie in the available dynamical
Lie algebra.  The Carlini formulation then leads to a constrained
brachistochrone equation, while the two-boundary algorithmic formulation
first identifies the boundary-induced direction and then asks how it can be
synthesized or approximated by the available gates.

The equivalence also fails in its simple form for anisotropic costs,
subspace endpoints, mixed states, open-system dynamics, or quantum
channels.  For higher-rank projectors, several principal angles appear
between the endpoint subspaces.  The commutator $[P_f,P_i]$ remains a
natural measure of boundary noncommutativity, but the exact geodesic
generator generally involves angle-dependent weights in the corresponding
two-dimensional principal-angle blocks.

\medskip
\noindent
The projector form clarifies the different roles of the two theories.
Carlini et al.'s brachistochrone is a theory of the fastest continuous quantum
motion under a specified Hamiltonian resource.  It is naturally suited to
quantum speed limits, pulse optimization, and physical gate-time bounds.

The two-boundary bridge formulation uses the same geodesic core as an
inverse-design principle for quantum computing.  A computational task
specifies an input boundary and an output condition.  These data first
define the ideal generator $\ii[P_f,P_i]$; the algorithmic question is then
how to realize that generator in a restricted model.  Oracle reflections in
Grover search, phase synthesis in the quantum Fourier transform, and
invariant two-dimensional rotations in quantum singular value
transformation can be viewed as model-dependent syntheses of such
boundary-induced rotations.

In the isotropic full-control limit the two approaches are dual
formulations of the same geodesic principle.  
Beyond that limit, Carlini's framework addresses constrained fastest physical motion, whereas the two-boundary formulation emphasizes the generator that a quantum circuit should synthesize from its computational boundary conditions.

\medskip
\noindent
We have rewritten the rank-one Carlini brachistochrone Hamiltonian in terms
of endpoint projectors and clarified its relation to the coherent
two-boundary bridge formulation.  The two theories coincide in their
geodesic core for pure states with full control and isotropic cost, with
$\omega=\Theta/T$ identifying the fixed-norm and fixed-time
normalizations.  The projector commutator form is especially useful for
quantum algorithms because it exposes the boundary-induced generator before
any particular gate decomposition is chosen.

The author was supported by the Cross-ministerial Strategic Innovation Promotion Program (SIP) of the Cabinet Office, Government of Japan (No. 23836436).

\label{lastpage}

\begin{thebibliography}{9}

\bibitem{Carlini2006}
A. Carlini, A. Hosoya, T. Koike, and Y. Okudaira:
Phys. Rev. Lett. \textbf{96} (2006) 060503.

\bibitem{Ohzeki2026}
M. Ohzeki:
arXiv:2607.10550.

\bibitem{Ohzeki2026-2}
M. Ohzeki and A. N. Jordan:
arXiv:2609.03097.

\bibitem{AnandanAharonov1990}
J. Anandan and Y. Aharonov:
Phys. Rev. Lett. \textbf{65} (1990) 1697.

\bibitem{Kato}
T. Kato:
J. Phys. Soc. Jpn. \textbf{5} (1950) 435.

\bibitem{Berry}
M. V. Berry:
J. Phys. A: Math. Theor. \textbf{42} (2009) 365303.

\end{thebibliography}
\end{document}